\documentclass[12pt]{elsarticle}

\usepackage{algorithm}
\usepackage{algorithmic}
\usepackage{booktabs}
\usepackage{multirow}
\usepackage[table, dvipsnames]{xcolor}
\usepackage{makecell}
\usepackage{amsmath}
\usepackage{pifont}
\usepackage{amssymb}
\usepackage{textcomp}
\usepackage{hyperref}

\usepackage{pgfplots}
\pgfplotsset{width=10cm,compat=1.9}

\usepackage{float}

\journal{Journal of Artificial Intelligence Research}

\begin{document}

\begin{frontmatter}

\title{Enhancement of Short Text Clustering by Iterative Classification}
\author{Md Rashadul Hasan Rakib, Norbert Zeh, Magdalena Jankowska, Evangelos Milios}
\address{Dalhousie University, Halifax, Nova Scotia, Canada\\ \{rakib, nzeh, jankowsk, eem\}@cs.dal.ca}









\begin{abstract}
Short text clustering is a challenging task due to the lack of signal contained in such short texts. 
In this work, we propose iterative classification as a method to boost the clustering quality (e.g., accuracy) of short texts.
Given a clustering of short texts obtained using an arbitrary clustering algorithm, iterative classification applies outlier removal to obtain outlier-free clusters. Then it trains a classification algorithm using the non-outliers based on their cluster distributions. Using the trained classification model, iterative classification reclassifies the outliers to obtain a new set of clusters. By repeating this several times, we obtain a much improved clustering of texts.
Our experimental results show that the proposed clustering enhancement method not only improves the clustering quality of different clustering methods (e.g., k-means, k-means\texttt{-{}-}, and hierarchical clustering) but also outperforms the state-of-the-art short text clustering methods on several short text datasets by a statistically significant margin.
\end{abstract}

\begin{keyword}
short text clustering, outlier removal, iterative classification.
\end{keyword}

\end{frontmatter}


\section{Introduction}
Due to technological advances, short texts are generated at large volumes from different sources, such as micro-blogging, question-answering, and social news aggregation websites. Organizing these texts (e.g., grouping them by topic) is an important step towards discovering trends (e.g., political, economic) in conversations~\cite{trendtweets:2010} and in other data mining tasks, such as data summarization~\cite{summarization:2019}, frequent pattern analysis~\cite{frequentpattern:2019}, and searching for and filtering information~\cite{IPMSearching:2017}. Clustering the texts into groups of similar texts is the foundation for many of these organization strategies~\cite{docClustering:2018, textclustDef:2017}. 
The objective of our research is to effectively remove the outliers from the clusters (obtained by a clustering algorithm) and then reassign them to the proper clusters so as to improve the clustering quality of different clustering methods as well as outperform the state-of-the-art short text clustering methods which are based on neural networks~\cite{XU201722, auto2019} and on a Dirichlet process multinomial mixture model~\cite{Yin:2016}.

The notion of our iterative classification is based on the idea of an iterative clustering algorithm called k-means\texttt{-{}-}~\cite{Shekhar2003kmeans__} which removes outliers in each iteration so as to obtain better clustering performance. Motivated by this, we remove outliers in each iteration of our iterative classification algorithm. Both k-means\texttt{-{}-} and our method compute representations of the outlier-free clusters. However, the way k-means\texttt{-{}-} and our method compute the representations are different. k-means\texttt{-{}-} represents each cluster by the average of the vectors of its non-outliers. In contrast, our method represents the clusters using a trained classification model. To obtain the model, it trains a classifier using the cluster labels of the non-outliers by following the notion of a previous work~\cite{labelbykmeans2004} which trained a classifier (i.e., support vector machine) using the labels obtained from k-means clustering. 
Using the trained model, iterative classification reclassifies the outliers to their target clusters. The resulting set of clusters obtained by our method is the input for the next iteration or, if this is the last iteration, the final set of clusters is returned by the algorithm.

The proposed clustering enhancement algorithm can be applied to any set of initial clusters and 
thus independent of the method used to obtain these clusters. The quality of the final set of clusters, however, does depend on the method used to compute the initial clusters. We use k-means~\cite{Vipin2000}, k-means\texttt{-{}-}~\cite{Shekhar2003kmeans__} and hierarchical clustering~\cite{Vipin2000} using dense and sparse similarity matrices to compute the initial clusters.
k-means and k-means\texttt{-{}-} clustering are applied to the vector representations of the texts. For hierarchical clustering, we use the text similarity matrix (dense or sparse). The dense similarity matrix stores the similarity value for each text pair, whereas the sparse similarity matrix keeps a certain number of most significant similarity values and discards the remaining ones (sets them to 0)~\cite{Gollub2010}.

The matrix sparsification can be performed using different criteria for choosing the values to discard. We consider two approaches here, one based on $k$-nearest neighbors~\cite{Vipin2000} and the other based on the similarity distribution\footnote{The similarity distribution based sparsification method has been described in our conference paper published in the ACM Symposium on Document Engineering, 2018.}~\cite{Rakib:2018}. The $k$-nearest neighbors method keeps the $k$ largest entries in each row.
In the similarity distribution based method, the number of similarities to keep in each row is not fixed. Instead, it is based on the distribution of the similarity values in each row, as characterized by he mean and standard deviation of these values. These sparsification methods are discussed in detail 
in Section~\ref{sec-sim-matrix-sparsification}. 

The two major contributions of this work are as follows:
\begin{itemize}
\item The proposed clustering enhancement method improves the clustering quality of various short text datasets and does not require human annotated data to train the classification model. The implementation of our clustering enhancement method and the datasets are publicly available for reproducibility\footnote{https://github.com/rashadulrakib/short-text-clustering-enhancement}.
\item The combination of hierarchical clustering (using a sparse similarity matrix based on similarity distribution~\cite{Rakib:2018}) and iterative classification performs better than other clustering methods combined with iterative classification. Moreover, this combination outperforms the state-of-the-art short text clustering methods by a statistically significant margin on the datasets with a moderate number of clusters.
\end{itemize}
The rest of this paper is organized as follows. A brief overview of related work is presented in Section~\ref{ic-relwork}. The proposed clustering enhancement method is described in Section~\ref{sec-ic}. 
We discuss similarity matrix sparsification methods
in Section~\ref{sec-sim-matrix-sparsification}. 
The 
short text clustering methods used in our experiments are described in Section~\ref{sec-clusteringalgos}.
The experimental results and evaluation of the proposed clustering enhancement method are
described in Section~\ref{sec-experiment}. We summarize the contributions and discuss future work in Section~\ref{sec-conclusion}.




\section{Related Work}\label{ic-relwork}

Our discussion of related work is divided into two parts.
First, we discuss related work on short text clustering. Then, we describe work on similarity matrix sparsification methods. 

\subsection{Short Text Clustering}
A major challenge in short text clustering is the sparseness of the vector representations of these texts resulting from the small number of words in each text.
Several clustering methods have been proposed in the literature to address this challenge, including methods based on text augmentation~\cite{Banerjee:2007, ZHENG20182444, semanticenrich:2019}, neural networks~\cite{XU201722, auto2019}, 
word co-occurrences~\cite{BTMXCheng2014, topicrepresentation:2019},
and Dirichlet mixture model~\cite{Yin:2016}.

A recent method based on text augmentation~\cite{ZHENG20182444} uses topic diffusion to augment each short text by finding words not appearing in the text that are related to its content. To find related words, this method determines possible topics for each text using the existing words. Then new words are added to each text; these new words are closely related to the text's topics based on the posterior probabilities of the new words given the words in the text. 
Another recent work~\cite{semanticenrich:2019} proposes two approaches to enrich short texts: one is based on distributional semantics, and the other is based on external knowledge resource. In the first approach, it generates embeddings of the words 
by training Word2Vec~\cite{w2vec2013} method on a raw corpus. Then, it finds the topmost similar word for each word in a text by computing similarity between the embedding of that word and the embeddings of other words and add the topmost similar words to the text. In the second approach, it uses an external knowledge resource called BabelNet~\cite{BabelNet:2012}, an integrated knowledge resource based on Wikipedia and WordNet.
Using BabelNet, it extracts synsets, categories, hypernyms, and glosses for each word in a text and adds them to the text as features.    
An earlier text augmentation method~\cite{Banerjee:2007} finds Wikipedia articles using the short text as query string and uses the articles' titles as features. 

A short text clustering method based on word embedding and a convolutional neural network called STC2-LE was proposed in~\cite{XU201722}. It uses a convolutional neural network to learn a text representation on which clustering is performed. Another short text clustering method based on weighted word embedding and autoencoder was proposed in~\cite{auto2019}. For each text, it calculates the average of the weighted pretrained word embeddings~\cite{w2vec2013} of its words. 
The weight of a word is calculated based on its inverse frequency in the corpus and on a hyperparameter~\cite{auto2019} which is then multiplied with its embedding to obtain weighted word embedding.
After that, the embeddings of the texts are feed into an autoencoder to obtain the low dimensional representation of the texts on which clustering is performed.
In general, neural network based approaches are computationally expensive. 


Biterm topic modeling (BTM)~\cite{BTMXCheng2014} is a topic modeling approach for short texts that learns topics from word co-occurrence patterns (i.e., biterms). Given a topic distribution produced by BTM for each text, clustering is performed by assigning a text to its most probable topic.
TRTD~\cite{topicrepresentation:2019} is a topic representative term discovery method for short texts that finds the groups of closely related topic representative terms based on the closeness and significance of the terms. The closeness is measured by the co-occurrence of the terms, and the significance is measured by their global occurrences throughout the entire text corpus. Based on the closeness and significance of the terms, TRTD discovers the term groups. Clustering is performed by assigning a text to a term group with which it shares the maximum number of terms.

A short text clustering method based on a Dirichlet process multinomial mixture model called GSDPMM was proposed in~\cite{Yin:2016}. GSDPMM does not partition the input into a pre-specified number of clusters. It processes the texts one by one and assigns each text to a new cluster or to one of the existing clusters based on two factors: a) a preference for a cluster with more texts and, b) a preference for a cluster whose texts share more words with the current text. Since GSDPMM does not know the number of clusters in advance, it uses a Dirichlet prior $\alpha$,
which corresponds to the prior probability of a text choosing a new cluster. The probability of assigning a document\footnote{In this paper, we interchangeably use the terms text and document.} to a new cluster becomes higher when $\alpha$ becomes larger, which in turn may lead to a larger than desired number of clusters. To overcome this issue, GSDPMM uses another Dirichlet prior, $\beta$, which corresponds to the prior probability of a text to choose a cluster that shares more similar content (i.e., common words) than other clusters with that text. The probability of choosing a new cluster for a document becomes smaller when $\beta$ becomes higher, which in turn prevents increasing the number of clusters. 
In general, a Dirichlet process mixture model based clustering algorithm requires tuning the parameters (i.e., $\alpha$ and $\beta$) to obtain the desired number of document clusters.

\subsection{Similarity Matrix Sparsification}

Sparsification of the text similarity matrix keeps the association between a text and its most similar (nearest) texts while breaking associations with less similar ones by setting the corresponding similarity scores to 0~\cite{Vipin2000}.
Several similarity matrix sparsification methods have been discussed in the literature, including ones based on a global threshold~\cite{Vipin2000}, nearest neighbors~\cite{Jain1988ACD46712}, and center vectors~\cite{Gollub2010}.  

Global threshold based similarity matrix sparsification is the simplest sparsification method. It removes all similarity values that are below a given threshold~\cite{Vipin2000}. The problem with this method is that some real clusters may be destroyed or merged because different clusters may have different similarity levels between the texts they contain. For example, the range of the similarity values between the texts in one cluster may be between 0.2 and 0.4 while the similarity values in another cluster may range from 0.5 to 0.8. If we set the global threshold to 0.5, then the similarity values in the first cluster are set to 0 and the cluster is destroyed.
A lower threshold, such as 0.15, may result in the inclusion of additional documents in the second cluster.

Nearest neighbors based methods for similarity matrix sparsification include $k$-nearest neighbors~\cite{Vipin2000}
and shared nearest neighbors~\cite{SharedNearestNeighbor:2018}. $k$-nearest neighbors sparsification keeps only the $k$ highest similarity scores for each text; 
the shared-nearest neighbors approach adds a condition that texts retaining similarity values with a particular text should share a prescribed number of neighbors. 

A similarity matrix sparsification method based on the center vector was proposed in~\cite{Gollub2010}. Texts are represented by \textit{tf}-\textit{idf} (term frequency-inverse document frequency) vectors and a center vector of the whole text collection is computed by averaging these vectors. The sparsification of the similarity matrix is performed by removing similarities between all pairs of texts that are not more similar to each other than the maximum similarities of these two texts to the center vector. 

\section{Enhancement of Clustering by Iterative Classification}\label{sec-ic}
Given a collection of short texts and a partition of these texts into clusters, 
iterative classification modifies the given cluster partition by detecting outliers in each cluster and changing the clusters to which they are assigned. This is repeated 
several times, hence the term iterative in the method's name. 
In each iteration, we generate training and test sets containing non-outliers and outliers respectively. 
Then we train a classification algorithm using the training set and classify the test set using the trained model. This iterative process repeats until the stopping criterion discussed in Section~\ref{subsub_ic_stop_criterion} is satisfied. The details are shown in Algorithm~\ref{algo:iterative_classification} and are described next.

\begin{algorithm}[H]
\caption{Enhancement of Clustering by Iterative Classification}
\label{algo:iterative_classification}
\begin{algorithmic}[1]
\REQUIRE 
\textit{D} = set of $n$ texts, \textit{L} = initial cluster labels of the texts in $D$, \textit{K} = number of clusters
\ENSURE Enhanced cluster labels of the texts
\STATE \textit{maxIteration} = 50
\STATE \textit{avgTextsPerCluster} = $n/K$
\FOR {$i=1$ to \textit{maxIteration}}
	\STATE Choose a parameter $P$ uniformly at random from the interval [$P_1$, $P_2$]. ($P_1$ and $P_2$ are parameters determined in Section~\ref{sub-sec-Experimental-Setup}. $P$ bounds the fraction of texts kept per cluster.)
	\STATE Remove outliers from each of the $K$ clusters defined by $L$ using an outlier detection algorithm. 
	\STATE If a cluster contains more than \textit{avgTextsPerCluster}$\times P$ texts, remove texts from that cluster uniformly at random so that exactly \\ \textit{avgTextsPerCluster}$\times P$ texts remain in the cluster.
	\STATE \textit{testSet} = texts removed in Steps 5 and 6 
	\\ \textit{trainingSet} = all the texts not in \textit{testSet}
	\STATE Train a classifier using the \textit{trainingSet} and classify the texts in \textit{testSet}. This assigns a new cluster label $L(t)$ to each text $t\in$ \textit{testSet}.
	\STATE Stop iterative classification if the per cluster text distribution becomes stable (as described in Section~\ref{subsub_ic_stop_criterion}).
	
\ENDFOR
\STATE return $L$

\end{algorithmic}
\label{algo-sparsify}
\end{algorithm}



\subsection{Training and Test Set Generation}
In each iteration, we choose a number $P$ that roughly corresponds to the fraction of texts selected for the training set. $P$ is chosen uniformly at random from an interval [$P_1$, $P_2$] determined in Section~\ref{sub-sec-Experimental-Setup}.

To generate the training set, we 
remove outliers from each of the $K$ clusters defined by the current cluster labels $L$.
To remove outliers, we use an outlier detection algorithm called Isolation Forest~\cite{Liu:2008:IF:1510528.1511387} which is applied to the \textit{tf}-\textit{idf} vector representations of the texts. The algorithm isolates each item based on the values of its features. At first, it selects a random feature of an item. Then it selects a random value between the minimum and maximum value of the selected feature.
After that it draws a straight line through that value to isolate the item. The algorithm repeats these steps until the item is isolated. The items which are isolated using fewer steps are considered as outliers as they exist in the low density region 
in the feature space~\cite{Liu:2008:IF:1510528.1511387}.
     
If after removing outliers, a cluster contains more than $\frac{n}{K}\times P$ texts, then we remove texts from that cluster uniformly at random
to reduce the number of texts in the cluster to $\frac{n}{K}\times P$. The reason of removing texts from each cluster is that we want each cluster to consist of roughly the same number of texts so as to reduce the bias of the classification algorithm. We add the removed texts to the test set and add the other texts to the training set.

\subsection{Training the Classification Model and Classification of the Test Set}
The texts in the training and test set are represented using the \textit{tf}-\textit{idf} vectors.
We train a classifier based on the clustering of the texts in the training set defined by their cluster labels. 
We train Multinomial Logistic Regression~\cite{surveyclassification:2012} using the training set to obtain a trained classifier.
Then we classify the texts in the test set using the trained classifier. 
This defines a new set of cluster labels of the texts in the test set and thus produces an updated cluster partition. 

\subsection{Stopping Criterion for Iterative Classification}\label{subsub_ic_stop_criterion}
Iterative classification stops when it reaches the maximum number of iterations (i.e., 50) or the sizes of the clusters become stable. Let $C_1,...,C_k$ and $C_1',...,C_k'$ be the clusters before and after an iteration, respectively. We consider the cluster sizes to be stable if $$\frac{1}{k}\sum_{i=1}^{k} ||C_i'|-|C_i||\leq 0.05\frac{n}{k}$$
For example, consider the problem of partitioning 100 texts into two clusters. Then the average cluster size is 50. If one iteration assigns 48 texts to the first cluster and 52 texts to the second cluster and the next iteration assigns 49 and 51 texts to these clusters, respectively, then the average absolute change of the cluster size is $\frac{1}{2}(|48-49|+|52-51|)=1$. Since this is less than 5\% of the average cluster size (50), 
we consider the cluster sizes to have stabilized. 

\section{Similarity Matrix Sparsification}\label{sec-sim-matrix-sparsification}
\subsection{k-Nearest Neighbors Sparsification}\label{subsec-knn-sparsification}
The $k$-nearest neighbors ($k$-NN) sparsification method~\cite{Jain1988ACD46712} uses the number of nearest neighbors $k$ as a parameter. A square $n \times n$ symmetric similarity matrix
$S=(s_{ij})$ is the input for $k$-NN sparsification method. The method criterion is to retain, for each text, exactly the $k$ highest similarities with this text outside of the diagonal.
For the text $t_{i}$, we retain a similarity ($s_{ij}$) between $t_{i}$ and other text $t_{j}$, if $s_{ij}$ is within the $k$ highest similarities of $t_{i}$. However, the similarity $s_{ji}$ between a text $t_{j}$ and other text $t_{i}$ may not be retained because $s_{ji}$ may not be within the $k$ highest similarities of $t_{j}$. Hence after applying this criterion, the resulting sparsified matrix can be a non-symmetric matrix. Therefore we symmetrize the sparsified similarity matrix by retaining both $s_{ij}$ and $s_{ji}$, if any of the similarities among $s_{ij}$ and $s_{ji}$ is retained in the sparsified similarity matrix.
\subsection{Similarity Distribution based Sparsification}\label{subsec-sim-distribution}
We propose a method to sparsify the similarity matrix of a set of objects based on the distribution of the similarity scores in this matrix. The sparsified matrix can then be used as an input to various similarity matrix-based clustering methods such as hierarchical clustering. Since our focus in this paper is on short text clustering, the objects in this paper are texts. 

Algorithm~\ref{algo:similarity_matrix_sparsification_SD} shows our sparsification method. The input is a symmetric similarity matrix for a set of $n$ texts. The goal is to increase the signal-to-noise ratio in the matrix by keeping only the ``most significant'' similarity scores and setting less significant similarity scores to 0.
Our criterion for setting entries to 0 may result in a non-symmetric matrix. Such a matrix requires symmetrization. We follow the ``sparsification with exclusion'' approach~\cite{Vipin2000} which sets an element $s_{ij}$ to zero only if the sparsification criterion retains neither $s_{ij}$ nor $s_{ji}$.

\begin{algorithm}[H]
\caption{Similarity Distribution Based Sparsification}
\label{algo:similarity_matrix_sparsification_SD}
\begin{algorithmic}[1]
\REQUIRE $S=(s_{ij})=n \times n$ similarity matrix, $K$ = number of clusters
\ENSURE $M=n \times n$ sparse similarity matrix
\STATE $M=(m_{ij})=n \times n$ identity matrix
\STATE $A=(a_{ij})=n \times n$ matrix, all elements are 0
\STATE $l = n/K-1$
\STATE $N=\left\lfloor\frac{n \times l}{2}\right\rfloor$, number of non-zero similarity scores to be retained above the diagonal
\FOR {$i=1$ to $n$}
	\STATE Calculate the mean $\mu_{i}$ and standard deviation $\sigma_{i}$ of all non-diagonal entries in row $i$. 
   	\FOR {$j=1$ to $n$ and $j \neq i$}
		\STATE $a_{ij} = (s_{ij}-\mu_{i})/\sigma_{i}$
	\ENDFOR
\ENDFOR
\STATE $L = \emptyset$ 
\FOR{$i=1$ to $n$}
\FOR {$j=i+1$ to $n$}
    \STATE Add the triplet ($i$,$j$, max($a_{ij},a_{ji}$)) to $L$
\ENDFOR
\ENDFOR
\STATE Sort the triplets in $L$ in descending order by their third components
\FOR {each of the first $N$ triplets ($i$,$j$, $\cdot$) in $L$} 
\STATE $m_{ij}=s_{ij}$
\STATE $m_{ji}=s_{ji}$
\ENDFOR
\STATE return $M$
\end{algorithmic}
\label{algo-sparsify}
\end{algorithm}


In contrast to the $k$-nearest neighbors method, the number of similarities to keep for each text is not fixed. Instead, it is based on the distribution of the similarity values between each text and all other texts. For each text $t_i$, we calculate the mean $\mu_i$ and standard deviation $\sigma_i$ of similarities between $t_i$ and all other texts. Then, we sparsify similarities between $t_i$ and other texts based on these statistics. In particular, we define the retaining criterion as follows: a similarity $s_{ij}$ is to be retained if and only if
\begin{equation}\label{eq-criterion}
s_{ij}>\mu_i + \alpha\sigma_{i},
\end{equation}
for some global factor $\alpha$. 
The factor $\alpha$ is chosen so that after applying the criterion and symmetrization of the matrix, the average number of non-zero elements outside of the diagonal per row is equal to $l=\frac{n}{K}-1$. Note that if each cluster has exactly $\frac{n}{K}$ elements and we return exactly the similarity scores between elements in the same cluster, then $l$ is the number of non-zero non-diagonal entries in each row. 

To choose the retained similarity values efficiently,
we use an auxiliary value
$a_{ij}=\frac{s_{ij} - \mu_i}{\sigma_{i}}$ for each similarity value $s_{ij}$. This is $s_{ij}$'s deviation from the mean of row $i$ normalized by the standard deviation of row $i$.
Our criterion from Eq.~\ref{eq-criterion} can then be restated as: a similarity $s_{ij}$ is to be retained if and only if $a_{ij}>\alpha$.
Since we follow the ``sparsification with exclusion'' approach for symmetrization, we keep $s_{ij}$ in the final symmetric matrix if the retaining criterion is fulfilled for $s_{ij}$ or for $s_{ji}$. Thus, if the average number of non-zero non-diagonal entries per row is to be $l$, we need to return $N=\left\lfloor\frac{n \times l}{2}\right\rfloor$ entries above the main diagonal, which is achieved by choosing $\alpha$ to be the $N^{th}$ largest value in $\{\max(a_{ij}, a_{ji})|1\leq i<j \leq n\}$.

Algorithm~\ref{algo:similarity_matrix_sparsification_SD} implements this strategy. In line 1, it initializes the sparsified matrix $M$ to be the identity matrix. Lines 5--10 calculate the auxiliary values $a_{ij}$ for all $i\neq j$. Lines 11--16 construct the list $L=\langle(i,j,\max(a_{ij}, a_{ji}))|1\leq i<j \leq n\rangle$. Line 17 sorts this list in descending order by the third components of its entries. By taking the first $N$ entries in this sorted list, we effectively choose $\alpha$ to be the $N^{th}$ largest value in $\{\max(a_{ij}, a_{ji})|1\leq i<j \leq n\}$ and select all pairs $(i,j)$ and $(j,i)$ such that $\max(a_{ij}, a_{ji})\geq\alpha$. Lines 18--21 copy the entries in these positions from $S$ to $M$. The result is the final sparse matrix, which we return in line 22.

\section{Methods for Clustering of Short Texts}\label{sec-clusteringalgos}
In this section, we discuss the methods used in our experiments for clustering of short texts. 
First, we discuss the k-means and k-means\texttt{-{}-} clustering algorithms. Then, we discuss hierarchical clustering using dense similarity matrix and hierarchical clustering using a sparse similarity matrix.

\subsection{k-means and k-means\texttt{-{}-}}
k-means clustering~\cite{kmeans:2010} is used to cluster a collection of short texts into $k$ clusters. Since k-means is highly sensitive to initial seed (i.e., cluster center) selection, we adopt a better seed selection approach called k-means++~\cite{kmeans_plus_Arthur:2007} to obtain better clustering performance. Using the initial cluster centers, k-means assigns each text to its closest center. Then the algorithm runs for a number of iterations. In each iteration, it recomputes the cluster centers using the texts assigned to each cluster and reassigns the texts to their closest centers. This iterative process continues until the algorithm reaches the maximum number of iterations or the cluster assignments becomes stable between two successive iterations. 

k-means\texttt{-{}-}~\cite{Shekhar2003kmeans__} is a variation of k-means clustering, in which outliers are removed in each iteration of the k-means clustering before recomputing the cluster centers.
To detect outliers, short texts are ranked in decreasing order using their distances to their nearest cluster centers and the $d$ (parameter for defining the total number of outliers) most distant texts are considered as outliers and removed from the clusters so that the cluster centers will become less sensitive to outliers. This has been confirmed to improve the clustering performance. 


\subsection{Hierarchical Clustering Using a Dense Similarity Matrix} 
Hierarchical agglomerative clustering uses a symmetric matrix storing pairwise similarities between documents. Such a matrix is dense if it stores a similarity between every pair of documents, the default. The clustering method starts with each document in its own clusters and repeatedly merges pairs of ``most similar'' clusters until only $k$ clusters remain ($k$ is the desired numbers of clusters, given as a parameter).

We used the fastcluster package as the clustering algorithm~\cite{JSSv053i09} in our experiments. This package supports many different measures between clusters to decide which cluster pairs to merge. The Ward linkage criterion produced by far the best results, so we used it in our experiments. The Ward linkage criterion assumes that the distances between data points are Euclidean distances and merges the two clusters that produce a merged cluster with minimum variance in each step. In our experiments, the distances between documents are cosine similarities, that is, non-Euclidean distances. The use of the Ward linkage criterion in our experiments can nevertheless be justified, and the effectiveness of this clustering method on our data sets explained. Indeed, for points close to each other on the unit space, the cosine of the angle they form with the origin is a close approximation of the distance between the two points on the space, which in turn closely approximates the Euclidean distance between the two points. If the number of clusters to be computed is not too small, then the points in each cluster are close to each other on the sphere, that is, the inter-point distances used to form these clusters have low distortion relative to their Euclidean distances. 

\subsection{Hierarchical Clustering Using a Sparse Similarity Matrix}
A dense similarity matrix provides the most detailed information about pairwise document similarities but the lowest similarity scores can be considered noise in the sense that they suggest (tenuous) connections between documents that are almost guaranteed to belong to different clusters. Setting these similarities to 0 increases the separation between clusters and produces better clustering results.

We consider two sparsification methods in our experiments: similarity distribution-based as discussed in Section~\ref{subsec-sim-distribution} and $k$-nearest neighbors discussed in Section~\ref{subsec-knn-sparsification}. We form clusters based on the two resulting sparse similarity matrices using the same hierarchical clustering method as in the previous section.

\section{Experiments}\label{sec-experiment}
\subsection{Datasets}
We used nine different datasets of short texts in our experiments. The basic properties of
these datasets 
are shown in Table~\ref{tbl_data_summary}.

\begin{table}[H]
\caption{Summary of the short text datasets}
\label{tbl_data_summary}
\begin{tabular}{|c|c|c|c|}
    \hline
Dataset & \#Clusters & \#Texts & Average \#words/text\\
\hline
SearchSnippet & 8 & 12340 & 17.03\\ 
\hline
SearchSnippet-test & 8 & 2280 & 17.18\\ 
\hline
AgNews & 4 & 8000 & 22.61\\
\hline
StackOverflow & 20 & 20000 & 8.23\\ 
\hline
Tweet & 89 & 2472 & 8.41\\ 
\hline
GoogleNews-TS
 & 152 & 11109 & 27.95\\ 
\hline
GoogleNews-T & 152 & 11109 & 6.14\\ 
\hline
GoogleNews-S & 152 & 11109 & 21.81\\ 
\hline
BioMedical & 20 & 20000 & 12.88\\ 
\hline

\end{tabular}
\end{table}

\textbf{SearchSnippet} is a dataset of search results from Google's search engine, containing 12340 snippets distributed into 8 groups~\cite{Phan:2008:LCS:1367497}. \textbf{SearchSnippet-test} is a subset of the SearchSnippet dataset 
consisting of 2280 search snippets distributed into 8 groups. \textbf{AgNews} is a subset of a dataset of news titles~\cite{zhang2015understanding}. It consists of 8000 texts in 4 topic categories (for each category, we randomly selected 2000 texts). \textbf{StackOverflow} is a subset of the challenge data published on Kaggle\footnote{https://www.kaggle.com/c/predict-closed-questions-on-stack-overflow/download/train.zip}, where 20000 question titles from 20 groups were randomly selected~\cite{XU201722}. \textbf{BioMedical} is a subset of the challenge data published on the BioASQ's website\footnote{http://participants-area.bioasq.org/}, where 20000 paper titles from 20 groups were randomly selected~\cite{XU201722}. 
The \textbf{Tweet} dataset consists of 2472 tweets grouped into 89 clusters~\cite{Yin:2016}. \textbf{GoogleNews-TS} contains titles and snippets of 11109 news articles related to 152 events~\cite{Yin:2016}. \textbf{GoogleNews-T} and \textbf{GoogleNews-S} were obtained from the GoogleNews-TS dataset by extracting only the title or only the text snippet of each news article.

\subsection{Experimental Setup}\label{sub-sec-Experimental-Setup}
\subsubsection{Experimental Setup for Iterative Classification}\label{sub-sec-Experimental-Setup}
We preprocessed the texts by removing stop words, punctuation and converting the texts to lowercase. Then we transformed each text into the \textit{tf}-\textit{idf} vector representation for a given text collection.
We considered various combinations of classification algorithms and outlier detection methods as the parameters of our clustering enhancement algorithm. The classification algorithms we considered were Multinomial Logistic Regression, SVM and k-Nearest Neighbour as described in~\cite{surveyclassification:2012}. The outlier detection methods we considered were One-Class SVM~\cite{Manevitz:2002:OSD:944790.944808} and Isolation Forest~\cite{Liu:2008:IF:1510528.1511387}. Among the various combinations of classification algorithms and outlier detection methods, Multinomial Logistic Regression with Isolation Forest performed best in preliminary experiments.
Thus we choose this combination in our experiments discussed below.

Each iteration of the iterative classification algorithm choose some percentage $P$ of each clusters as the training set and reassigns the remaining documents to clusters based on a classifier trained using this training set; $P$ is chosen uniformly at the random from some interval [$P_1$, $P_2$]. To justify this approach and to determine optimal choices for $P_1$ and $P_2$, we ran preliminary experiments discussed in detail in~\ref{app_exp1exp2} using two representative datasets (SearchSnippet-test and Tweet). Specifically, we considered choosing $P$ uniformly at random from the interval [$P_1$, $P_2$] or choosing a fixed percentage $P$ in every iteration. For the former method, we determined the optimal combination of $P_1$ and $P_2$ ($P_1=0.5$ and $P_2=0.95$). For the later, we determined the optimal choice of $P$ ($P=0.6$). 
Choosing $P$ uniformly at random from the interval [0.5, 0.95]
resulted in cluster accuracies of 82.21 and 87.70, respectively, for the two representative datasets. 
Choosing a fixed percentage $P=0.6$ in every iteration resulted in cluster accuracies of 80.25 and 84.88, respectively. Thus we chose $P_1=0.5$ and $P_2=0.95$ and chose $P$ uniformly at random from this interval in all remaining experiments.

\subsubsection{Experimental Setup for Clustering}\label{expsettingcluster}
To perform the clustering of short texts, we used the preprocessed texts described in Section~\ref{sub-sec-Experimental-Setup}. After that, texts were represented as vectors using pretrained word embeddings. The pretrained word embeddings used in our work were Glove (Global Vectors for Word Representation)~\cite{pennington2014glove} and BioASQ~\cite{bioasq2018}. The Glove embedding\footnote{http://nlp.stanford.edu/data/glove.42B.300d.zip} was trained using the Glove method~\cite{pennington2014glove} on Wikipedia dumps. The BioASQ embedding\footnote{bioasq.lip6.fr/tools/BioASQword2vec/} was trained using the Word2Vec method~\cite{w2vec2013} on abstracts of biomedical publications. We used the Glove embedding for all datasets except the biomedical dataset since these datasets contained terms related to general domains such as news and search snippets.
For the biomedical dataset, the BioASQ embedding was more appropriate due to its specific focus on biomedical terms.


To cluster a collection of texts, we represented each text by the average of the vectors of all words in the text. 
Then, we applied the five different clustering methods
described in Section~\ref{sec-clusteringalgos} to the text vectors. For the k-means and k-means\texttt{-{}-}~\cite{Shekhar2003kmeans__} clustering algorithms, we used the text vectors as the points to be clustered. For hierarchical clustering, we constructed the dense similarity matrix by computing similarities between the vectors using cosine similarity for all the text pairs. 
After that, we sparsified the dense similarity matrix using the $k$-NN and similarity distribution-based ($SD$) sparsification methods. Then we applied hierarchical agglomerative clustering using dense (HAC) and sparse similarity matrices (HAC$\_k$-NN and HAC$\_SD$).





\subsection{Results}
First, we describe the experimental results of short text clustering.
Then, we discuss the impact of iterative classification to improve the initial clustering of short texts. After that, we compare the results obtained using iterative classification with the results of state-of-the-art short text clustering methods. Finally, we discuss whether the improvement obtained using iterative classification is statistically significant.

\subsubsection{Description of Experimental Results}
In our experiments, we use nine datasets of short texts. Among them five datasets consit of smaller number of clusters which are SearchSnippet, SearchSnippet-test, AgNews, StackOverflow, and BioMedical. The other four datasets consist of larger number of clusters which are Tweet, GoogleNews-TS, GoogleNews-T, and GoogleNews-S. We used accuracy (ACC) and normalized mutual information (NMI) as the evaluation measures for different clustering algorithms (as in \cite{XU201722}).
The clustering results (ACC, NMI) of the datasets having smaller number of clusters are shown in Tables~\ref{tbl_hac_kmeans_state_acc_main_body_stc2} and~\ref{tbl_hac_kmeans_state_nmi_main_body_stc2}. The last three rows of Tables~\ref{tbl_hac_kmeans_state_acc_main_body_stc2} and~\ref{tbl_hac_kmeans_state_nmi_main_body_stc2} show the ACC and NMI scores obtained using the state-of-the-art short text clustering methods, STC2-LE~\cite{XU201722}, SIF-Auto~\cite{auto2019}, and GSDPMM~\cite{Yin:2016}. The clustering results (ACC, NMI) of the datasets having larger number of clusters are shown in Tables~\ref{tbl_hac_kmeans_state_acc_main_body_gsdpmm} and~\ref{tbl_hac_kmeans_state_nmi_main_body_gsdpmm}. The last row of Tables~\ref{tbl_hac_kmeans_state_acc_main_body_gsdpmm} and~\ref{tbl_hac_kmeans_state_nmi_main_body_gsdpmm} show the ACC and NMI scores obtained using the state-of-the-art short text clustering method GSDPMM~\cite{Yin:2016}. The ACC and NMI scores of five clustering algorithms both before and after iterative classification for the nine datasets are shown in these four Tables. The results with or without the $\_IC$ suffix are the results with or without iterative classification. The best result (ACC, NMI) for each dataset is shown in bold.

\begin{table}[H]
\centering
\caption{ACC of different clustering methods, 
their corresponding enhancements by iterative classification, and state-of-the-art methods 
for short text clustering. $^\ding{1}$ indicates that this method is statistically significantly inferior to its corresponding enhancement by iterative classification (in terms of ACC). * indicates that this method is statistically significantly inferior to HAC$\_SD\_IC$ (in terms of ACC).
The best ACC scores are shown in bold.}
\small
\label{tbl_hac_kmeans_state_acc_main_body_stc2}

\renewcommand{\tabcolsep}{1pt}

\begin{tabular}{|c|c|c|c|c|c|}
\hline
 & \multicolumn{5}{c|} {Datasets} \\ \cline{2-6}
Clustering & Search & Search & AgNews & Stack & Bio \\
Methods&Snippet&SnippetTest& &Overflow &Medical\\ \cline{2-6}
&ACC($\%$)&ACC($\%$)&ACC($\%$)&ACC($\%$)&ACC($\%$)\\ \hline

HAC$\_SD$& 82.69$^\ding{1}$ &89.47$^\ding{1}$& 81.84$^\ding{1}$ &64.80$^\ding{1}$ & 40.13$^\ding{1}$ \\ 
\rowcolor{lightgray}
HAC$\_SD\_IC$& \textbf{87.67}$\pm$0.63 & \textbf{92.16}$\pm$0.85 & \textbf{84.52}$\pm$0.50 &\textbf{78.73}$\pm$0.17 & 47.78$\pm$0.51 \\ \Xhline{2\arrayrulewidth}

HAC$\_k$-NN& 79.08$^\ding{1}$* &87.14$^\ding{1}$* & 76.83$^\ding{1}$* &58.11$^\ding{1}$* & 39.75$^\ding{1}$* \\
HAC$\_k$-NN$\_IC$&83.19*$\pm$0.61 &90.76*$\pm$1.79& 81.83*$\pm$0.35 &70.07*$\pm$0.11 & 46.17*$\pm$1.10 \\ \Xhline{2\arrayrulewidth}

HAC& 76.54$^\ding{1}$* &77.06$^\ding{1}$* & 76.56$^\ding{1}$* &61.64$^\ding{1}$* & 38.86$^\ding{1}$* \\ 
HAC$\_IC$& 80.63*$\pm$0.69 &83.92*$\pm$2.66& 81.13*$\pm$1.22 &67.69*$\pm$2.12 & 46.13*$\pm$0.92 \\ \Xhline{2\arrayrulewidth}

k-Means& 63.89$^\ding{1}$*$\pm$1.15&63.22$^\ding{1}$*$\pm$1.79& 58.17$^\ding{1}$*$\pm$1.87 &41.54$^\ding{1}$*$\pm$2.16 & 36.92$^\ding{1}$*$\pm$0.81 \\ 
k-Means$\_IC$& 83.13*$\pm$0.69 &82.84*$\pm$2.32& 78.06*$\pm$3.13 &69.89*$\pm$1.52 & 43.50*$\pm$1.38 \\ \Xhline{2\arrayrulewidth}

k-means\texttt{-{}-}& 47.42$^\ding{1}$*$\pm$1.13 &61.96$^\ding{1}$*$\pm$1.98& 62.48$^\ding{1}$*$\pm$2.13 &43.77$^\ding{1}$*$\pm$0.39 & 39.95$^\ding{1}$*$\pm$1.21 \\ 
k-means\texttt{-{}-}$\_IC$& 79.77*$\pm$2.67 &75.29*$\pm$2.79& 77.45*$\pm$3.49 &69.25*$\pm$1.88 & 45.61*$\pm$3.19 \\ \Xhline{2\arrayrulewidth}

STC2-LE& 78.29*$\pm$2.72 & &  &53.81*$\pm$3.37 & 44.81*$\pm$1.72 \\ \hline
SIF-Auto & 79.13*$\pm$1.27 & &  &59.85*$\pm$1.81 & \textbf{55.73}$\pm$1.97 \\ \hline
GSDPMM& 38.67*$\pm$2.78 &50.91*$\pm$2.19& 39.53*$\pm$1.89 &29.36*$\pm$1.47 & 28.09*$\pm$1.81 \\ \hline
\end{tabular}
\end{table}

\begin{table}[H]
\centering
\caption{NMI of different clustering methods, 
their corresponding enhancements by iterative classification, and state-of-the-art methods 
for short text clustering. $^\ding{1}$ indicates that this method is statistically significantly inferior to its corresponding enhancement by iterative classification (in terms of NMI). * indicates that this method is statistically significantly inferior to HAC$\_SD\_IC$ (in terms of NMI).
The best NMI scores are shown in bold.
}
\small
\label{tbl_hac_kmeans_state_nmi_main_body_stc2}

\renewcommand{\tabcolsep}{1pt}
\begin{tabular}{|c|c|c|c|c|c|}
\hline
 & \multicolumn{5}{c|} {Datasets} \\ \cline{2-6}
Clustering & Search & Search & AgNews & Stack & Bio \\
Methods&Snippet&SnippetTest& &Overflow &Medical\\ \cline{2-6}
&NMI($\%$)&NMI($\%$)&NMI($\%$)&NMI($\%$)&NMI($\%$)\\ \hline

HAC$\_SD$& 63.76$^\ding{1}$ &78.73$^\ding{1}$ & 54.57$^\ding{1}$ &59.48$^\ding{1}$ & 33.51$^\ding{1}$ \\ 
\rowcolor{lightgray}
HAC$\_SD\_IC$& \textbf{71.93}$\pm$1.04 &\textbf{85.55}$\pm$1.09&\textbf{59.07}$\pm$0.84 &\textbf{73.44}$\pm$0.35 & 41.27$\pm$0.36 \\ \Xhline{2\arrayrulewidth}

HAC$\_k$-NN& 60.51$^\ding{1}$* &76.42$^\ding{1}$*& 52.43$^\ding{1}$* &54.06$^\ding{1}$* & 32.19$^\ding{1}$* \\
HAC$\_k$-NN$\_IC$&65.49*$\pm$0.97 &83.17*$\pm$1.17& 56.02*$\pm$0.86 &68.88*$\pm$0.43 & 38.78*$\pm$0.53 \\ \Xhline{2\arrayrulewidth}

HAC& 59.41$^\ding{1}$* &70.99$^\ding{1}$* & 52.82$^\ding{1}$* &54.46$^\ding{1}$* & 31.01$^\ding{1}$* \\ 
HAC$\_IC$& 63.61*$\pm$1.09 &77.49*$\pm$1.11& 56.57*$\pm$1.23 &61.76*$\pm$1.35 & 38.50*$\pm$0.61 \\ \Xhline{2\arrayrulewidth}

k-Means& 43.75$^\ding{1}$*$\pm$1.31 &51.54$^\ding{1}$*$\pm$0.92& 35.26$^\ding{1}$*$\pm$2.01 &38.01$^\ding{1}$*$\pm$2.12 & 33.71$^\ding{1}$*$\pm$0.29 \\ 
k-Means$\_IC$& 66.27$^{\ding{1}}\pm$1.00 &76.88$^{\ding{1}}\pm$2.64& 52.32$^{\ding{1}}\pm$2.47 &69.84$^{\ding{1}}\pm$0.66 & 38.08$^{\ding{1}}\pm$0.81 \\ \Xhline{2\arrayrulewidth}

k-means\texttt{-{}-}& 47.43$^\ding{1}$*$\pm$1.65 &49.73$^\ding{1}$*$\pm$2.15& 39.68$^\ding{1}$*$\pm$1.15 &41.89$^\ding{1}$*$\pm$0.86 & 34.49*$\pm$1.93 \\ 
k-means\texttt{-{}-}$\_IC$& 63.01*$\pm$1.69 &71.11*$\pm$2.40& 51.05*$\pm$3.63 &69.64*$\pm$1.28 & 35.63*$\pm$2.82 \\ \Xhline{2\arrayrulewidth}

STC2-LE& 64.72*$\pm$1.37 & &  &49.51*$\pm$1.63& 38.42*$\pm$0.87 \\ \hline
SIF-Auto& 57.72*$\pm$1.43 & &  &55.59*$\pm$1.23& \textbf{47.21}$\pm$1.19 \\ \hline
GSDPMM& 40.57*$\pm$1.86 &48.96*$\pm$2.38& 42.83*$\pm$2.14 &30.62*$\pm$1.13 & 32.04*$\pm$1.58 \\ \hline
\end{tabular}
\end{table}


\begin{table}[H]
\centering
\caption{ACC of different clustering methods, 
their corresponding enhancements by iterative classification, and state-of-the-art methods 
for short text clustering. $^\ding{1}$ indicates that this method is statistically significantly inferior to its corresponding enhancement by iterative classification (in terms of ACC). * indicates that this method is statistically significantly inferior to HAC$\_SD\_IC$ (in terms of ACC).
The best ACC scores are shown in bold.}
\small
\label{tbl_hac_kmeans_state_acc_main_body_gsdpmm}

\renewcommand{\tabcolsep}{1pt}

\begin{tabular}{|c|c|c|c|c|}
\hline
 & \multicolumn{4}{c|} {Datasets} \\ \cline{2-5}
Clustering &  Tweet & Google & Google & Google  \\
Methods& &News-TS&News-T&News-S\\ \cline{2-5}
&ACC($\%$)&ACC($\%$)&ACC($\%$)&ACC($\%$)\\ \hline

HAC$\_SD$ &89.62$^\ding{1}$ &85.76$^\ding{1}$ &81.75$^\ding{1}$ &80.63$^\ding{1}$ \\ 
\rowcolor{lightgray}
HAC$\_SD\_IC$& \textbf{91.52}$\pm$0.99 &92.25$\pm$0.10 &\textbf{87.18}$\pm$0.21 &\textbf{89.02}$\pm$0.12 \\ \Xhline{2\arrayrulewidth}

HAC$\_k$-NN&88.14$^\ding{1}$* &85.56$^\ding{1}$* &82.53$^\ding{1}$* &78.77$^\ding{1}$* \\
HAC$\_k$-NN$\_IC$&90.09*$\pm$0.38 &91.24*$\pm$0.15 &85.86*$\pm$0.31 &86.99*$\pm$0.34 \\ \Xhline{2\arrayrulewidth}

HAC& 87.94$^\ding{1}$* &88.61$^\ding{1}$* &81.54$^\ding{1}$* &79.03$^\ding{1}$* \\ 
HAC$\_IC$&91.45$\pm$0.28 &\textbf{93.56}$\pm$0.27 &85.52*$\pm$0.33 &88.33$\pm$0.18 \\ \Xhline{2\arrayrulewidth}

k-Means& 82.85$^\ding{1}$*$\pm$0.24 &83.63$^\ding{1}$*$\pm$0.76 &78.15$^\ding{1}$*$\pm$0.53 &77.23$^\ding{1}$*$\pm$0.66 \\ 
k-Means$\_IC$ & 86.24*$\pm$1.27 & 85.52*$\pm$1.34&84.46*$\pm$0.62 &84.88*$\pm$0.69 \\ \Xhline{2\arrayrulewidth}

k-means\texttt{-{}-} &84.18$^\ding{1}$*$\pm$0.19 &80.07$^\ding{1}$*$\pm$0.88 &71.83$^\ding{1}$*$\pm$0.76 &72.18$^\ding{1}$*$\pm$1.11 \\ 
k-means\texttt{-{}-}$\_IC$&87.67*$\pm$1.35 &83.77*$\pm$1.14 &81.70*$\pm$0.71 &82.65*$\pm$0.77 \\ \Xhline{2\arrayrulewidth}

GSDPMM& 84.39*$\pm$2.19 &79.51*$\pm$2.73 &74.88*$\pm$1.93 &71.04*$\pm$2.16 \\ \hline
\end{tabular}
\end{table}

\begin{table}[H]
\centering
\caption{NMI of different clustering methods, 
their corresponding enhancements by iterative classification, and state-of-the-art methods 
for short text clustering. $^\ding{1}$ indicates that this method is statistically significantly inferior to its corresponding enhancement by iterative classification  (in terms of NMI). * indicates that this method is statistically significantly inferior to HAC$\_SD\_IC$ (in terms of NMI).
The best NMI scores are shown in bold.}
\small
\label{tbl_hac_kmeans_state_nmi_main_body_gsdpmm}
\renewcommand{\tabcolsep}{1pt}
\begin{tabular}{|c|c|c|c|c|}
\hline
 & \multicolumn{4}{c|} {Datasets} \\ \cline{2-5}
Clustering &  Tweet & Google & Google & Google  \\
Methods& &News-TS&News-T&News-S\\ \cline{2-5}
&NMI($\%$)&NMI($\%$)&NMI($\%$)&NMI($\%$)\\ \hline

HAC$\_SD$ &85.19 &87.97$^\ding{1}$ &84.17$^\ding{1}$ &83.51$^\ding{1}$ \\ 
\rowcolor{lightgray}
HAC$\_SD\_IC$ &\textbf{86.87}$\pm$0.13 &93.21$\pm$0.14 &\textbf{87.87}$\pm$1.00 &\textbf{89.96}$\pm$0.11 \\ \Xhline{2\arrayrulewidth}

HAC$\_k$-NN&84.65$^\ding{1}$* &88.51$^\ding{1}$* &85.62* &82.69$^\ding{1}$* \\
HAC$\_k$-NN$\_IC$&86.02$\pm$0.25 &93.34$\pm$0.14 &86.93$\pm$0.16 &87.71*$\pm$0.16 \\ \Xhline{2\arrayrulewidth}

HAC &83.87$^\ding{1}$* &90.32$^\ding{1}$* &85.27* &82.26$^\ding{1}$* \\ 
HAC$\_IC$&86.35$\pm$0.28 &\textbf{94.40}$\pm$0.11 &86.76$\pm$0.17 &89.85$\pm$0.18 \\ \Xhline{2\arrayrulewidth}

k-Means&79.86$^\ding{1}$*$\pm$0.25 &86.65$^\ding{1}$*$\pm$0.83 &81.52$^\ding{1}$*$\pm$0.41 &80.88$^\ding{1}$*$\pm$0.84 \\ 
k-Means$\_IC$ &83.46*$\pm$1.01 &91.97*$\pm$0.41 &86.45*$\pm$0.30 &87.10*$\pm$0.28  \\ \Xhline{2\arrayrulewidth}

k-means\texttt{-{}-}  &80.37$^\ding{1}$*$\pm$0.79 &83.72$^\ding{1}$*$\pm$1.34 &70.78$^\ding{1}$*$\pm$0.81 &78.46$^\ding{1}$*$\pm$1.63 \\ 
k-means\texttt{-{}-}$\_IC$&84.51*$\pm$1.21 &92.11*$\pm$0.40 &85.70*$\pm$0.34 &86.42*$\pm$0.43 \\ \Xhline{2\arrayrulewidth}

GSDPMM &86.13$\pm$1.91 &91.92*$\pm$2.25 &86.19*$\pm$2.07 &86.69*$\pm$2.21 \\ \hline
\end{tabular}
\end{table}

To compensate for the dependence of k-Means, k-Means\texttt{-{}-} on the choice of cluster seeds, we ran the k-Means and k-Means\texttt{-{}-} clustering algorithms 20 times on the same dataset and performed iterative classification on the clustering obtained in each run. After that, we calculated the mean and standard deviation of the 20 clustering results (ACC, NMI) obtained by k-Means, k-means\texttt{-{}-}, k-Means$\_IC$ and k-means\texttt{-{}-}$\_IC$ for each dataset.

We ran hierarchical agglomerative clustering (HAC), HAC$\_k$-NN, and HAC$\_SD$ 
only once since 
HAC is deterministic. 
However, the enhancement of the clustering obtained by iterative classification varies between runs since the training and test sets are chosen randomly in each iteration. Therefore, we ran the iterative classification 20 times on the clustering obtained using HAC, HAC$\_k$-NN and HAC$\_SD$, and again calculated the mean and standard deviation of each of the 20 clustering results obtained using HAC$\_IC$, HAC$\_k$-NN$\_IC$ and HAC$\_SD\_IC$ for each dataset. 



\subsubsection{Impact of Iterative Classification}
We evaluated whether iterative classification improves the initial clustering obtained using different clustering algorithms. We consider iterative classification to improve the clustering for a given dataset if both ACC and NMI are increased using iterative classification.


Tables~\ref{tbl_hac_kmeans_state_acc_main_body_stc2}, \ref{tbl_hac_kmeans_state_nmi_main_body_stc2}, \ref{tbl_hac_kmeans_state_acc_main_body_gsdpmm}, and~\ref{tbl_hac_kmeans_state_nmi_main_body_gsdpmm} show that iterative classification improves the initial clustering of short texts in terms of both ACC and NMI.
In our experiments, iterative classification improved the ACC and NMI of k-means clustering by 10\% and 11\%, respectively on average for the 9 datasets. Likewise, iterative classification improved the ACC of k-means\texttt{-{}-}, HAC, HAC$\_k$-NN and HAC$\_SD$ by 12\%, 4\%, 5\% and 5\%, respectively on average. 
The NMI of k-means\texttt{-{}-}, HAC, HAC$\_k$-NN and HAC$\_SD$ improved by 12\%, 5\%, 5\% and 6\%, respectively on average for these 9 datasets.

For most of the datasets (except GoogleNews-TS) the best clustering result (ACC, NMI) was obtained by applying iterative classification to the clustering obtained by hierarchical agglomerative clustering using SD sparsification (HAC$\_SD$).
The reason is that HAC$\_SD$~\cite{Rakib:2018} produces better initial clustering than other clustering methods for these datasets and the enhancement of clustering depends on the initial clustering.

\subsubsection{Comparison with State-of-the-Art Methods}
Our second comparison aims to assess how the results of iterative classification in conjunction with the different clustering methods compare to state-of-the-art short text clustering methods,
specifically STC2-LE~\cite{XU201722}, SIF-Auto~\cite{auto2019}, and GSDPMM~\cite{Yin:2016}.

STC2-LE is a short text clustering method based on a convolutional neural network (CNN). Tables~\ref{tbl_hac_kmeans_state_acc_main_body_stc2} and~\ref{tbl_hac_kmeans_state_nmi_main_body_stc2} show that 
HAC$\_SD$ and HAC$\_k$-NN  
outperform STC2-LE\footnote{We were unable to reproduce the clustering for other short text datasets using STC2-LE and SIF-Auto.} for the SearchSnippet, StackOverflow and BioMedical datasets in terms of ACC and NMI.

SIF-Auto is a short text clustering method based on an autoencoder. Tables~\ref{tbl_hac_kmeans_state_acc_main_body_stc2} and~\ref{tbl_hac_kmeans_state_nmi_main_body_stc2} show that 
HAC$\_SD$, HAC$\_k$-NN, HAC$\_SD$, k-Means$\_IC$, and k-means\texttt{-{}-}$\_IC$ 
outperform SIF-Auto for the SearchSnippet, and StackOverflow datasets in terms of ACC and NMI. However, on the Biomedical dataset, the performance of SIF-Auto is better than any clustering method and its corresponding enhancement by iterative classification.

GSDPMM is a short text clustering method based on a Dirichlet process multinomial mixture model and does not require the number of clusters to be produced as a parameter. GSDPMM has a tendency to produce many clusters.
Therefore, it performs better for datasets that have many clusters. 
Tables~\ref{tbl_hac_kmeans_state_acc_main_body_gsdpmm} and~\ref{tbl_hac_kmeans_state_nmi_main_body_gsdpmm} show that the clustering results (ACC, NMI) of GSDPMM are better for the Tweet, GoogleNews-TS, GoogleNews-T, GoogleNews-S datasets than for the SearchSnippet, SearchSnippetTest, AgNews, StackOverflow and Biomedical datasets. The Tweet, GoogleNews-TS, GoogleNews-T, GoogleNews-S datasets have many more clusters than the other datasets.

For datasets with many clusters, the various clustering methods with iterative classification are competitive with GSDPMM. For datasets with fewer clusters, the clustering methods based on the iterative classification perform significantly better than GSDPMM and these are the datasets where GSDPMM struggles.

\subsubsection{Statistical Significance Testing of Clustering Performance}
Our third comparison aims to investigate whether the clustering improvements achieved by iterative classification are statistically significant. In particular, we perform two investigations: a) whether the improved results achieved by iterative classification are statistically significant with the results of their corresponding clustering methods. b) whether the improved results achieved by our best clustering method HAC$\_SD\_IC$ are statistically significant with the results of different clustering methods (with or without iterative classification and state-of-the-art methods).
For significance testing, we performed a two-tailed paired t-test (with significance level $\alpha=0.05$) using the pairwise differences of clustering results (ACC, NMI) of 20 runs obtained by different pairs of clustering methods. $^\ding{1}$ in the Tables indicate that the clustering method is statistically significantly inferior to its corresponding enhancement by iterative classification in terms of ACC and NMI (i.e., the improvement achieved by iterative classification is statistically significant with its corresponding clustering method). * in the Tables indicate that the clustering method is statistically significantly inferior to our best clustering method HAC$\_SD\_IC$ in terms of ACC and NMI (i.e., the improvement achieved by HAC$\_SD\_IC$ is statistically significant with that particular clustering method). 

For example, it is shown in Table~\ref{tbl_hac_kmeans_state_nmi_main_body_stc2} that iterative classification improves the accuracy of HAC from 76.54 to 80.63 for SearchSnippet dataset and this improvement 
is statistically significant at 0.05 level. Therefore, the accuracy 76.54 is marked with a $^\ding{1}$. In the same Table, the clustering accuracy 76.54 obtained by HAC is statistically significantly inferior to the accuracy 87.67 obtained by HAC$\_SD\_IC$ for the SearchSnippet dataset. Therefore, the accuracy 76.54 is also marked with a *.

The ACC and NMI scores achieved by the state-of-the-art methods STC2-LE, SIF-Auto, and GSDPMM are statistically significantly inferior to the ACC and NMI scores achieved by our best clustering method HAC$\_SD\_IC$ for the datasets SearchSnippet, StackOverflow, GoogleNews-TS, GoogleNews-T, and GoogleNews-S as shown in Tables~\ref{tbl_hac_kmeans_state_acc_main_body_stc2}, ~\ref{tbl_hac_kmeans_state_nmi_main_body_stc2},~\ref{tbl_hac_kmeans_state_acc_main_body_gsdpmm}, and~\ref{tbl_hac_kmeans_state_nmi_main_body_gsdpmm}. Therefore those scores are marked with *. For the BioMedical dataset, the ACC and NMI scores achieved by HAC$\_SD\_IC$ are statistically significantly better than that of STC2-LE and GSDPMM. However, SIF-Auto performs better than HAC$\_SD\_IC$ on the BioMedical dataset. On the Tweet dataset, HAC$\_SD\_IC$ achieved better ACC and NMI scores than GSDPMM. The improvement in ACC score is statistically significant, but for NMI, the improvement is not statistically significant.

\section{Conclusion and Future Work}\label{sec-conclusion}
We have demonstrated that 
iterative classification enhances the clustering of short texts for various short text datasets based on initial clusters obtained using
such as k-means, k-means\texttt{-{}-}, hierarchical agglomerative clustering (HAC), HAC using $k$-NN and $SD$ sparsification methods. The most promising results were obtained by applying iterative classification to the clustering obtained by HAC using the proposed $SD$ sparsification (denoted by HAC$\_SD\_IC$). Experimental results show that HAC$\_SD\_IC$ outperforms a state-of-the-art neural network based short text clustering method (STC2-LE) in terms of both ACC and NMI. 
Moreover, HAC$\_SD\_IC$ outperforms other state-of-the-art short text clustering methods which are based on autoencoder (SIF-Auto) and on Dirichlet process multinomial mixture model (GSDPMM), in terms of ACC and NMI on several short text datasets. 
The proposed clustering enhancement method advances the state of the art in short text clustering, which is important in the following practical contexts such as social media monitoring, product recommendation, and customer feedback analysis. The proposed method is a generic clustering enhancement approach for short texts where various classification algorithms, initial clustering and number of clusters can be easily integrated.

In the future, we will apply our clustering enhancement algorithm to long documents to investigate whether iterative classification leads to performance improvements.
We also plan to use phrase similarity as a basis for computing text similarity so as to obtain better text similarity scores, since the performance of clustering algorithms depends on the quality of individual text similarity scores.

\section{References}
\bibliography{mybibfile}

\pagebreak
\appendix
\section{Parameters Tuning of Iterative Classification}\label{app_exp1exp2}
In this appendix, we discuss the experiments we performed to tune the performance of iterative classification. Specifically, \ref{appndx_exp1} determines the optimal percentage of the dataset to be chosen as the training set in each iteration if we choose a fixed percentage in each iteration. \ref{appndx_exp2} then determines the optimal lower bound $P_1$ and upper bound $P_2$ when choosing the percentage of data to include in the training set uniformly at random from the interval [$P_1$, $P_2$] in each iteration. By comparing the results obtained by these two variants of iterative classification using their optimal parameters, we conclude that choosing a random percentage from the interval [$P_1$, $P_2$] in each iteration produces better results.

We used two representative datasets (SearchSnippet-test, Tweet) in our experiments. SearchSnippet-test is representative of the datasets SearchSnippet, Agnews, StackOverflow and BioMedical since SearchSnippet-test and these four datasets contain a moderate number of clusters.
Tweet is representative of the datasets GoogleNews-TS, GoogleNews-T and GoogleNews-S since Tweet and these three datasets contain a larger number of clusters.


\subsection{Fixed-Size Training Set}\label{appndx_exp1} 
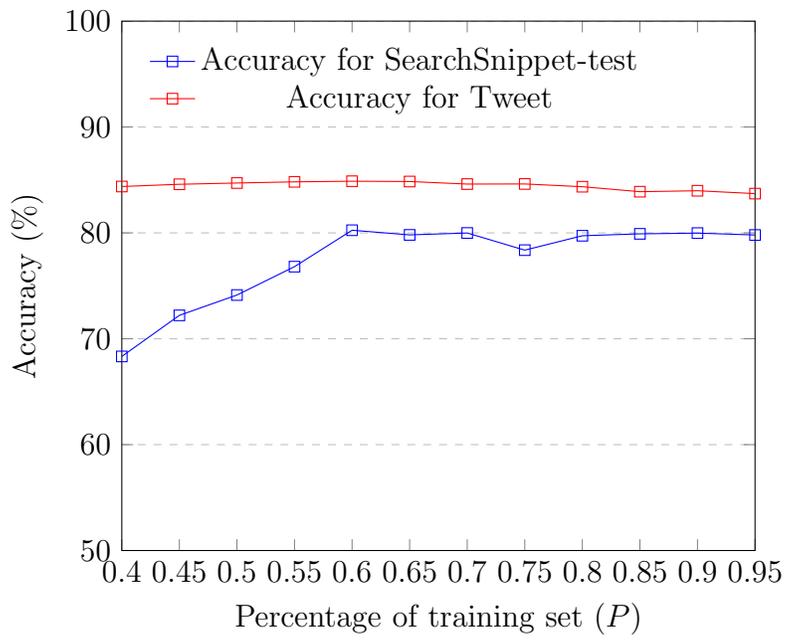
\begin{figure}
\begin{tikzpicture} 
\begin{axis}[
    title={},
    xlabel={Percentage of training set ($P$)},
    ylabel={Accuracy (\%)},
    xmin=0.4, xmax=0.95,
    ymin=50, ymax=100,
    xtick={0.4,0.45, 0.5, 0.55, 0.6, 0.65, 0.7, 0.75, 0.8, 0.85, 0.9, 0.95},
    ytick={50, 60, 70, 80, 90, 100},
    legend pos=north west,
    ymajorgrids=true,
    grid style=dashed,
    legend style={draw=none}
]
 
\addplot[
    color=blue,
    mark=square,
    ]
    coordinates {
    (0.4,68.33)(0.45,72.21)(0.50,74.13)(0.55,76.81)(0.60,80.25)(0.65,79.81)(0.70,79.99)(0.75,78.37)(0.80,79.73)(0.85,79.91)(0.90,79.98)(0.95,79.80)
    };
     \addlegendentry{Accuracy for SearchSnippet-test}
\addplot[
    color=red,
    mark=square,
    ]
     coordinates {
    (0.4,84.38)(0.45,84.59)(0.50,84.71)(0.55,84.82)(0.60,84.88)(0.65,84.85)(0.70,84.61)(0.75,84.62)(0.80,84.36)(0.85,83.89)(0.90,83.98)(0.95,83.71)
    };
    \addlegendentry{Accuracy for Tweet}

\end{axis}
\end{tikzpicture}
\caption{Accuracy of the clustering enhancement algorithm using various fixed percentage of training sets (in each iteration) for the datasets SearchSnippet-test and Tweet. Clustering enhancement is performed on the initial clustering obtained by k-means.} \label{figure_fixed_p_tweet_searchsnippet_test}
\end{figure}

In this experiment, we used a fixed percentage $P$ of the input as training set
in each iteration of our clustering enhancement algorithm. We tested values of $P$ between 0.40 and 0.95 in increment of 0.05.
We ran the algorithm on the initial clustering of the datasets SearchSnippet-test and Tweet obtained using k-means.
Figure~\ref{figure_fixed_p_tweet_searchsnippet_test} shows the accuracies achieved on the two representative datasets for the different choices of $P$. 
Based on these results, we observe that the best clustering accuracies of 80.25 and 84.88 obtained for SearchSnippet-test and Tweet, respectively, were obtained for $P$=0.60.
\subsection{Experiment 2: Variable-Size Training Set}\label{appndx_exp2}
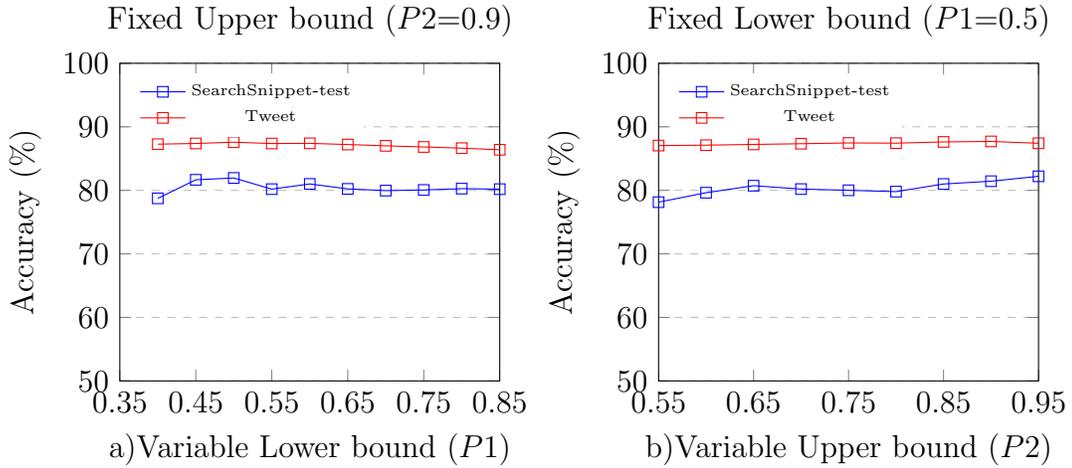
\begin{figure}[h]  
\centering 

\begin{tikzpicture}   
   \begin{axis}[
    title={Fixed Upper bound ($P2$=0.9)},
    xlabel={a)Variable Lower bound ($P1$)},
    ylabel={Accuracy (\%)},
    xmin=0.35, xmax=0.85,
    ymin=50, ymax=100,
    xtick={0.35, 0.45, 0.55, 0.65, 0.75, 0.85},
    ytick={50, 60, 70, 80, 90, 100},
    legend pos=north west,
    ymajorgrids=true,
    grid style=dashed,
    legend style={draw=none, font=\tiny},
    scale = .60
]
 
\addplot[
    color=blue,
    mark=square,
    ]
    coordinates {
    (0.4,78.74)(0.45,81.66)(0.50,81.95)(0.55,80.19)(0.60,81.01)(0.65,80.23)(0.70,79.96)(0.75,80.06)(0.80,80.27)(0.85,80.19)
    };
     \addlegendentry{SearchSnippet-test}
\addplot[
    color=red,
    mark=square,
    ]
     coordinates {
    (0.4,87.27)(0.45,87.38)(0.50,87.58)(0.55,87.37)(0.60,87.40)(0.65,87.21)(0.70,87.01)(0.75,86.83)(0.80,86.64)(0.85,86.39)
    };
    \addlegendentry{Tweet}

\end{axis}
\end{tikzpicture}
\begin{tikzpicture}  
   \begin{axis}[
    title={Fixed Lower bound ($P1$=0.5)},
    xlabel={b)Variable Upper bound ($P2$)},
    ylabel={Accuracy (\%)},
    xmin=0.55, xmax=0.95,
    ymin=50, ymax=100,
    xtick={0.55,0.65, 0.75, 0.85, 0.95},
    ytick={50, 60, 70, 80, 90, 100},
    legend pos=north west,
    ymajorgrids=true,
    grid style=dashed,
    legend style={draw=none, font=\tiny},
    scale = .60
]
 
\addplot[
    color=blue,
    mark=square,
    ]
    coordinates {
    (0.55,78.14)(0.60,79.63)(0.65,80.73)(0.70,80.20)(0.75,80.01)(0.80,79.80)(0.85,81.00)(0.90,81.44)(0.95,82.21)
    };
     \addlegendentry{SearchSnippet-test}
\addplot[
    color=red,
    mark=square,
    ]
     coordinates {
    (0.55,87.05)(0.60,87.11)(0.65,87.23)(0.70,87.34)(0.75,87.46)(0.80,87.43)(0.85,87.62)(0.90,87.70)(0.95,87.41)
    };
    \addlegendentry{Tweet}

\end{axis}
\end{tikzpicture} 

\caption{Accuracy of the clustering enhancement  algorithm using random percentage of training sets (in each iteration) between $P1$ and $P2$ for the datasets SearchSnippet-test and Tweet. In a) we vary $P1$ and keep $P2$ fixed. In b) we vary $P2$ and keep $P1$ fixed. Initial clustering is obtained by k-means.} \label{fig:2_1_and_2_2}  
\end{figure}

For these experiments, it would have been ideal to try all possible combinations of $0.4\leq P_1<P_2\leq0.95$ in increments of 0.05, similar to the experiments in the previous section. This is very time-consuming. To limit the time we spent on these experiments, we instead treated $P_1$ and $P_2$ as independent. In a first set of experiments, we varied $P_1$ from 0.4 to 0.85 in increments of 0.05 while keeping $P_2$ fixed at 0.90. As Figure~\ref{fig:2_1_and_2_2} (left) shows, the best accuracies of 81.95 and 87.58 for the two datasets were obtained for $P_1=0.5$. In a second set of experiments, we fixed $P_1$ at this value and let $P_2$ vary from 0.55 to 0.95 in increments of 0.05. The best accuracies of 82.21 and 87.70 we obtained for $P_2=0.95$ and $P_2=0.9$ for the SearchSnippet-test and Tweet datasets, respectively. Since the performance on the SearchSnippet-test dataset is more sensitive to the choice of $P_2$, we chose $P_1=0.5$ and $P_2=0.95$ as the final parameter combination.

\end{document}